\documentstyle[art12,12pt,epsf]{article}
\def\zbbar{$Z^0\rightarrow b \overline{b}$}
\def\zcbar{$Z^0\rightarrow c \overline{c}$}

\begin{document}

\renewcommand{\thefootnote}{\fnsymbol{footnote}}

\begin{titlepage}
  \begin{flushright}
SLAC-PUB-8201\\
 July, 1999\\
  \end{flushright}

\vskip 1.5cm

\begin{center}

\Large{\bf  New Measurement of $A_b$ at the $Z^0$ Resonance
using a Vertex-Charge Technique.\footnote{Work supported in part by the
Department of Energy contract  DE--AC03--76SF00515.}}
\vskip 0.3cm

The SLD Collaboration$^{**}$


\end{center}

\vglue 2.0cm
\begin{abstract}
\noindent

We present a new preliminary measurement of the parity-violation parameter
$A_b$
using a self-calibrating vertex-charge technique.
In the SLD experiment we observe hadronic decays
of $Z^0$ bosons produced in collisions between longitudinally
polarized electrons and unpolarized positrons at the SLAC Linear
Collider.  
A sample of $b\bar{b}$ events is selected
using the topologically reconstructed mass of $B$ hadrons.
From our 1996--1998 data sample of 
approximately 400,000 hadronic $Z^0$ decays, we obtain
$A_b=0.897 \pm 0.027 (\mbox{stat})^{+0.036}_{-0.034} (\mbox{syst})$.

\end{abstract}

\vglue 3.cm
\centerline{\it Paper Contributed to the International Europhysics Conference on High Energy}
\centerline{\it Physics (EPS-HEP 99), July 15-21, 1999, Tampere, Finland, Ref 6-473;}
\centerline{\it and the XIX International Symposium on Lepton and Photon Interactions }
\centerline{\it  at High Energies, August 9-14, 1999, Stanford, CA, USA}

\end{titlepage}

Measurements of $b$ quark production asymmetries at the $Z^0$ pole 
determine the extent of parity violation in the $Zb\bar{b}$ coupling.
At Born level, the differential cross section for 
the process $e^+e^-\rightarrow Z^0 \rightarrow b\bar{b}$
can be expressed 
as a function of the polar angle  
$\theta$ of the $b$ quark
relative to the electron beam direction, 
\begin{equation}
\label{DIFFXSECT}
\sigma^b(\xi) \equiv
d\sigma_b/d\xi 
 \propto (1 - A_e P_e)(1 + \xi^2) + 
2 A_b (A_e - P_e) \xi ,\nonumber
\end{equation}
where
$P_e$ is the longitudinal polarization of the electron beam, $\xi=\cos\theta$.
The parameters $A_f = 2v_f a_f/(v_f^2+a_f^2)$, ($f=e$ or $b$) where 
$v_f\ (a_f)$ is the vector (axial vector) coupling of the fermion $f$
to the $Z^0$ boson,
express the extent of parity violation in the $Zf\bar{f}$ coupling.

From the conventional forward-backward asymmetries formed with an 
unpolarized electron beam
($P_e = 0$),
such as used by the LEP experiments,
only the product of parity-violaton parameters $A_eA_b$ 
can be measured ~\cite{LEPEWWG}.
For a polarized electron beam,
it is possible to measure $A_b$ directly by forming the left-right
forward-backward asymmetry 
\cite{BLRV88}
\begin{equation}
\label{ALRFB}
 \tilde{A_b} \equiv {A_{LRFB}}^b(\xi) = {[\sigma_L^b(\xi) - \sigma_L^b(-\xi)] -
              [\sigma_R^b(\xi) - \sigma_R^b(-\xi)] \over
              \sigma_L^b(\xi) + \sigma_L^b(-\xi) +
              \sigma_R^b(\xi) + \sigma_R^b(-\xi)} =
  |P_e| A_b {2 \xi \over 1 + \xi^2 }\ , 
\end{equation}
where $L,R$ refers to \zbbar\ decays produced with a predominantly
left-handed (negative helicity) or right-handed (positive
helicity) electron beam, respectively. 
The measurement of the double asymmetry
eliminates the dependence on the initial state coupling.
The quantity $\tilde{A_b}$ is largely independent of
propagator effects that modify the effective weak mixing angle
and thus is complementary to other electroweak
asymmetry measurements performed at the $Z^0$ pole.
 
In this paper we present a preliminary direct measurement of $A_b$ 
from data collected in the SLC Large Detector (SLD)
during its 1996--1998 run. 
We use an inclusive vertex mass tag to select a
sample of \zbbar\ events, and use the charge of the reconstructed
secondary vertex to identify the sign of the
charge of the underlying quark. To measure the accuracy
of the quark charge assignment,  
we use a simple self-calibration technique which greatly 
reduces the model dependence of the result.
The result from this analysis is competitive with and 
complementary to our previous measurements using jet-charge \cite{JETQ98},
lepton \cite{LTAG98} and $K^{\pm}$ tags\cite{KTAG98}.

A detailed description of the SLD 
can be found elsewhere \cite{RBPRD}.
Charged particles are tracked
in the Central Drift Chamber (CDC) in a
uniform axial magnetic field of 0.6T.
In addition, new a pixel-based CCD
vertex detector (VXD3), installed in 1996,
provides an accurate measure
of particle trajectories close to the beam axis.
Recent improvements in the charged particle tracking
algorithm have further improved the overall tracking performance.
The measured $r\phi$ ($rz$)
track impact parameter resolution now approaches
$9 \mu$m ($11 \mu$m) for high momentum tracks, 
while multiple scattering contributions are
$33 \,\mu$m~$/(p_{\perp}\,{\rm sin}^{3/2}\theta)$ in both projections
($z$ is the
coordinate parallel to the beam axis and
$p_{\perp}$ is the momentum
in GeV/c perpendicular to the beamline).
The momentum resolution of the combined SLD tracking systems is
$(\delta p _{\perp} / p _{\perp})^2 = (.01)^2 + (.0026p_{\perp})^2$.
The thrust axis is reconstructed using the liquid argon
calorimeter, which covers a range of $|\cos \theta| < 0.98$.
The uncertainty in the position of 
the primary vertex ($PV$) is $\sim 4~\mu$m
transverse to the beam axis and $\sim 20~\mu$m (for $b\bar{b}$
events) along the beam axis.

Events are classified as hadronic  $Z^0$ decays if
they contain: (1) at least seven well-measured tracks 
(as described in Ref.~\cite{RBPRD}),
(2) a visible charged energy of at least 20 GeV,
and (3) have a thrust axis polar angle satisfying
$|\cos\theta_{thrust}| < 0.7$. The resulting hadronic sample 
from the $1996-98$ data consists of $\sim 400,000$ events
with a non-hadronic background estimated to be $< 0.1 \%$.
Events classified as having more than
three jets by
the  JADE jet-finding
algorithm with $y_{cut} = 0.02$ \cite{JADE}, 
using reconstructed charged tracks as input,
are discarded.

To increase the \zbbar\ content of the sample, 
we select events with
reconstructed secondary decay vertices\cite{RB};
the inclusive vertexing procedure is based on a
3-dimensional topological algorithm\cite{ZVTOP}.
We calculate the invariant mass of the reconstructed vertex ($M_{vtx}$),
correcting for missing transverse
momentum to partially account for neutral particles and tracking
inefficiencies. 
We require that the event contain
two vertices well separated from the interaction point, with least one
vertex (the ``tag'' vertex) with $M_{vtx} > 2.0 GeV/c^2$. 
This results in a sample of 24112 candidate \zbbar\
decays. 
The $b$-hadron purity and efficiency of this 
selection are calculated from the data by counting single-
and double-tagged events, assuming the Standard Model values
for the \zbbar\ and \zcbar\ fractions $R_b$ and $R_c$\cite{RBPRD},
and using the Monte Carlo (MC) simulation to predict the 
charm-hadron efficiency $\epsilon_c$.

The quark and/or antiquark direction is determined by the charge
of the reconstructed vertex: e.g., reconstructing a vertex with
$Q_{vtx} = +1$ in a given hemisphere indicates the $\overline{b}$
quark was produced in that hemisphere. To improve the accuracy of
the vertex charge reconstruction, additional quality tracks which 
were not used in the original topological vertex finding are ``attached''
to the toplogical vertex if they pass a set of criteria
determined from MC simulation to select primarily $B$ and $D$ decay
tracks. These attached tracks are used to improve the vertex charge
reconstruction only. All reconstructed vertices with a net charge
$Q_{vtx} \neq 0$ are used in this analysis; in the cases where
two charged vertices are reconstructed in a single event, the event
is discarded from further analysis if both vertices have the same sign. 
The MC simulation predicts that a reconstructed 
$b$-hadron vertex with $M_{vtx} > 2.0$ GeV/c$^2$ correctly assigns the 
underlying quark charge 
with an average probability $<p_b> = (73.0 \pm 0.2)\%$.

The value of $A_b$ is extracted  via a fit to a
maximum likelihood function
based on the differential
cross-section (see Eq. \ref{DIFFXSECT}), which provides
a somewhat more efficient estimate of $A_b$ than the
simple left-right forward-backward asymmetry of Eq. \ref{ALRFB}:
\begin{eqnarray}
\label{LIKELIHOODFUNCT}
 \rho^i (A_b) = (1-A_eP_e^i)(1+(T_z^i)^2)
 + 2(A_e-P_e^i)T_z^i
  [A_bf^i_b(2p_{b}^i-1)&\!\!\!\!\!(1-\Delta_{QCD,b}^i) + \nonumber\\
A_cf^i_c(2p_{c}^i-1)(1-\Delta_{QCD,c}^i) +
A_{bckg}(1-f^i_b-f^i_c)(2p_{bckg}^i-1)],&
\end{eqnarray}
where $P_e^i$ is the signed polarization of the electron beam for 
event $i$, $f_{b(c)}^i$ 
the probability
that the event is a $Z^0\rightarrow b{\overline b}(c{\overline c})$
decay, parametrized as a function of
the secondary vertex mass,
and $\Delta_{QCD,b,c}^i$ are final-state
QCD corrections, to be discussed later.
$A_{bckg}$ is the estimated 
asymmetry of residual
$u\overline u$, $d\overline d$, and $s \overline s$ final states.
The parameters $p_{ }$ are estimates of the probability
that the sign of $Q_{vtx}$ accurately reflects the charge of the
respective underlying quark, and are functions of the 
secondary vertex mass.

In order to reduce dependence on $B$ decay and detector reconstruction 
modelling, 
we use a self-calibrating technique to measure $p_{b}$ directly from 
the data.
Defining $N_{++}$ ($N_{+-}$) as the number of events with
two reconstructed vertices of the same (opposite) sign,
one can solve for $p_{b}$:
\begin{equation}
{p_{b}} = {1\over{2}}(\sqrt{{N_{+-} - N_{++}}\over{N_{+-} + N_{++}}} + 1)
\end{equation}
where we have assumed both vertices have the same correct-sign
probability $p_{b}$. In general this is not the case, so we
use the MC to determine the mass-dependent shape $p_{b}(M_{vtx})$
and correct the above equation appropriately. Uncertainties in 
the mass-dependence of $p_{b}$ are included in our systematic error
estimate (see below).
When applied to our MC simulation, this self-calibration
technique gives an average correct-sign probability 
$<p_b(M_{vtx}>2.0)> = (73.6 \pm 0.5)\%$, in good agreement with the MC 
``true'' value quoted above. The error here is due only to the 
limited statistics of the self-calibration technique.  
The same technique applied to the data yields
$<p_b(M_{vtx}>2.0)> = (75.6 \pm 0.9)\%$, and we use this value in
the analysis. The MC mass dependence $p_{b}(M_{vtx})$ is used
to extrapolate this value to other masses.

Final-state gluon radiation reduces the observed asymmetry
from its Born-level value. 
 This effect is incorporated in our analysis by applying a correction
$\Delta_{QCD}(|\cos\theta|)$
to the maximum likelihood function 
(Eq.~\ref{LIKELIHOODFUNCT}).
This correction is based on the ${\Large \scriptstyle{O}}(\alpha_s)$ 
calculation for massive
final state quarks of Stav and Olsen ~\cite{STAVOLSEN},
which ranges from 
$\Delta^{SO}_{QCD}(\left |\cos\theta \right |) \sim 0.05$ 
at $|\cos\theta| = 0$ 
to $\sim 0.01$ at $|\cos\theta| = 1$.
However, QCD radiative effects are mitigated
by the use of the thrust axis to estimate the $b$-quark direction, the
\zbbar\ enrichment algorithm, the self-calibration procedure,
and the cut on the number of jets. 
A MC simulation of the analysis chain indicates that these effects can
be represented by a $\cos\theta$-independent suppression factor,
$x_{QCD}=0.50 \pm 0.25$, such that
$\Delta_{QCD} =  x_{QCD} \Delta^{SO}_{QCD}$.
The effects of ${\scriptstyle{\Large O}}(\alpha_s^2)$ QCD radiation \cite{HOQCD},
which are dominated by gluon splitting to $b\bar{b}$,
lead to an additional correction $\delta A_b/A_b = +0.004 \pm 0.002$.


The dependence of the $b$-tagging efficiency
upon the secondary vertex mass is taken from
the simulation, with the overall tagging efficiency
derived from the single- and double-tagging rates \cite{RB} observed in the data.
Tagging efficiencies for charm and $uds$ events are estimated using
the MC simulation, as is the
charm correct-signing probability $p_{c}$.  The value of
$A_c$ is set to its Standard Model value of 0.67, and the value
of $A_{bckg}$ is set to zero.  
After a small (+0.2\%) correction \cite{ZFITTER} for initial state
radiation and $Z$-$\gamma$ interference,
the value of $A_b$ extracted from the fit
is $A_b=0.897\pm0.027\,\,(stat)$.
This result is found to be insensitive to the value of the 
$b$-tag mass cut.

  We have investigated a number of systematic effects which can change the
measured value of $A_b$; these are summarized in Table~\ref{SYSERR}.
The uncertainty in $p_b$ due to the
statistical uncertainties in the data self-calibration technique
corresponds to a $+3.4/-3.2\%$ uncertainty in $A_b$\footnote{The error in the self-calibrated $p_b$ is symmetric, but the 
corresponding error in the event weight (the ``analyzing power'', 
$= 2p_b - 1$) is asymmetric.}.
We have estimated the effects of possible biases in the
self-calibration technique by comparing the MC
true value of $p_b(M_{vtx})$ with the 
self-calibrated value of the same quantity determined using
the same MC as a trial dataset. We observe no bias, and
assign a  $1.0$\% systematic uncertainty
in $A_b$ due to our limited MC statistics. 
The uncertainty in the MC modelling of the 
$M_{vtx}$ dependence of $p_b$ is included in the 
tracking efficiency corrections (see below).
In addition, while the mean value of the self-calibration
parameter $p_b$ is constrained by the data, it 
has a $\cos\theta$ dependence due to the fall-off of the
tracking efficiency at high $\cos\theta$ which must be
estimated using the simulation, leading to a $0.6$\% uncertainty
in $A_b$. We also rely on the MC to correctly model the
vertex charge distribution of the light-flavor background
(dominantly \zcbar\ ) which is subtracted from the raw 
counts $N_{++}$ and $N_{+-}$; we conservatively take
a $\pm 50\%$ relative uncertainty on this subtraction,
which results in a $0.1$\% uncertainty
in $A_b$.

The extracted value of $A_b$ is sensitive to our estimate
of the  
$Z^0 \rightarrow c {\overline c}$ background, which tends to reduce
the observed asymmetry due to the positive charge of the underlying
$c$ quark. 
The uncertainty in the purity estimate of $\Pi_b =  (98.6 \pm 0.6)\%$
is dominated by the uncertainties in the charm tagging efficiency
($\epsilon_c = 0.009 \pm 0.001$) 
and leads to
a $0.9\%$ uncertainty in $A_b$. Details of the estimate of the
light and charmed quark efficiencies can be found in Ref.~\cite{RB}.

In addition, agreement between the data and MC simulation charged 
track multiplicity distributions is obtained only after the inclusion 
of additional ad-hoc tracking inefficiency. This random inefficiency was
parametrized as a function of total track momentum, and averages 0.5 
charged tracks per event.
However, without this correction applied, the MC correct-sign probability
is $<p_b(M_{vtx}>2.0)> = (74.4 \pm 0.3)\%$, in better agreement with the data than
the lower value obtained with the correction turned on.
Moreover, the agreement in the data/MC $M_{vtx}$ spectra is somewhat
better without the ad-hoc correction applied. 
Completely removing this
additional correction from the MC results in a $1.0\%$ change
in $A_b$, which is also included as a systematic error.
The uncertainty in the beam polarization $P_e$ is taken
from a preliminary estimate performed for the SLD $A_{LR}$
analysis\cite{PETER}. 

Combining all systematic uncertainties in quadrature yields
a total relative systematic uncertainty of $+4.0/-3.8\%$.

In conclusion, we have exploited the highly polarized SLC electron
beam to perform a direct measurement of
\begin{equation}
  A_b=0.897\pm 0.027(\mbox{stat})^{+0.036}_{-0.034}(\mbox{syst}).
\end{equation}

We thank the personnel of the SLAC accelerator department and the
technical staffs of our collaborating institutions for their
outstanding efforts on our behalf.
This work was supported by the U.S. Department of Energy 
and National Science Foundation, 
the UK Particle Physics and Astronomy Research Council,
the Istituto Nazionale di Fisica Nucleare of Italy,
the Japan-US Cooperative Research Project on High Energy Physics,
and the Korea Science and Engineering Foundation. 

\newpage
\begin{center}
{\Large \bf The SLD Collaboration}
\def\iADEL{$^{(1)}$}
\def\iAOMORI{$^{(2)}$}
\def\iBOLO{$^{(3)}$}
\def\iBRI{$^{(4)}$}
\def\iBRUN{$^{(5)}$}
\def\iBU{$^{(6)}$}
\def\iCINC{$^{(7)}$}
\def\iCOLO{$^{(8)}$}
\def\iCOLU{$^{(9)}$}
\def\iCSU{$^{(10)}$}
\def\iFERR{$^{(11)}$}
\def\iFRAS{$^{(12)}$}
\def\iILLI{$^{(13)}$}
\def\iJHU{$^{(14)}$}
\def\iLBL{$^{(15)}$}
\def\iLTU{$^{(16)}$}
\def\iMASS{$^{(17)}$}
\def\iMISSI{$^{(18)}$}
\def\iMIT{$^{(19)}$}
\def\iMOSCOW{$^{(20)}$}
\def\iNAGO{$^{(21)}$}
\def\iOREG{$^{(22)}$}
\def\iOXF{$^{(23)}$}
\def\iPADO{$^{(24)}$}
\def\iPERU{$^{(25)}$}
\def\iPISA{$^{(26)}$}
\def\iRAL{$^{(27)}$}
\def\iRUTG{$^{(28)}$}
\def\iSLAC{$^{(29)}$}
\def\iSOGA{$^{(30)}$}
\def\iSOONG{$^{(31)}$}
\def\iTENN{$^{(32)}$}
\def\iTOHO{$^{(33)}$}
\def\iUCSB{$^{(34)}$}
\def\iUCSC{$^{(35)}$}
\def\iUVIC{$^{(36)}$}
\def\iVAND{$^{(37)}$}
\def\iWASH{$^{(38)}$}
\def\iWISC{$^{(39)}$}
\def\iYALE{$^{(40)}$}

  \baselineskip=.75\baselineskip  
\mbox{Kenji  Abe\unskip,\iNAGO}
\mbox{Koya Abe\unskip,\iTOHO}
\mbox{T. Abe\unskip,\iSLAC}
\mbox{I. Adam\unskip,\iSLAC}
\mbox{T.  Akagi\unskip,\iSLAC}
\mbox{H. Akimoto\unskip,\iSLAC}
\mbox{N.J. Allen\unskip,\iBRUN}
\mbox{W.W. Ash\unskip,\iSLAC}
\mbox{D. Aston\unskip,\iSLAC}
\mbox{K.G. Baird\unskip,\iMASS}
\mbox{C. Baltay\unskip,\iYALE}
\mbox{H.R. Band\unskip,\iWISC}
\mbox{M.B. Barakat\unskip,\iLTU}
\mbox{O. Bardon\unskip,\iMIT}
\mbox{T.L. Barklow\unskip,\iSLAC}
\mbox{G.L. Bashindzhagyan\unskip,\iMOSCOW}
\mbox{J.M. Bauer\unskip,\iMISSI}
\mbox{G. Bellodi\unskip,\iOXF}
\mbox{A.C. Benvenuti\unskip,\iBOLO}
\mbox{G.M. Bilei\unskip,\iPERU}
\mbox{D. Bisello\unskip,\iPADO}
\mbox{G. Blaylock\unskip,\iMASS}
\mbox{J.R. Bogart\unskip,\iSLAC}
\mbox{G.R. Bower\unskip,\iSLAC}
\mbox{J.E. Brau\unskip,\iOREG}
\mbox{M. Breidenbach\unskip,\iSLAC}
\mbox{W.M. Bugg\unskip,\iTENN}
\mbox{D. Burke\unskip,\iSLAC}
\mbox{T.H. Burnett\unskip,\iWASH}
\mbox{P.N. Burrows\unskip,\iOXF}
\mbox{R.M. Byrne\unskip,\iMIT}
\mbox{A. Calcaterra\unskip,\iFRAS}
\mbox{D. Calloway\unskip,\iSLAC}
\mbox{B. Camanzi\unskip,\iFERR}
\mbox{M. Carpinelli\unskip,\iPISA}
\mbox{R. Cassell\unskip,\iSLAC}
\mbox{R. Castaldi\unskip,\iPISA}
\mbox{A. Castro\unskip,\iPADO}
\mbox{M. Cavalli-Sforza\unskip,\iUCSC}
\mbox{A. Chou\unskip,\iSLAC}
\mbox{E. Church\unskip,\iWASH}
\mbox{H.O. Cohn\unskip,\iTENN}
\mbox{J.A. Coller\unskip,\iBU}
\mbox{M.R. Convery\unskip,\iSLAC}
\mbox{V. Cook\unskip,\iWASH}
\mbox{R.F. Cowan\unskip,\iMIT}
\mbox{D.G. Coyne\unskip,\iUCSC}
\mbox{G. Crawford\unskip,\iSLAC}
\mbox{C.J.S. Damerell\unskip,\iRAL}
\mbox{M.N. Danielson\unskip,\iCOLO}
\mbox{M. Daoudi\unskip,\iSLAC}
\mbox{N. de Groot\unskip,\iBRI}
\mbox{R. Dell'Orso\unskip,\iPERU}
\mbox{P.J. Dervan\unskip,\iBRUN}
\mbox{R. de Sangro\unskip,\iFRAS}
\mbox{M. Dima\unskip,\iCSU}
\mbox{D.N. Dong\unskip,\iMIT}
\mbox{M. Doser\unskip,\iSLAC}
\mbox{R. Dubois\unskip,\iSLAC}
\mbox{B.I. Eisenstein\unskip,\iILLI}
\mbox{I.Erofeeva\unskip,\iMOSCOW}
\mbox{V. Eschenburg\unskip,\iMISSI}
\mbox{E. Etzion\unskip,\iWISC}
\mbox{S. Fahey\unskip,\iCOLO}
\mbox{D. Falciai\unskip,\iFRAS}
\mbox{C. Fan\unskip,\iCOLO}
\mbox{J.P. Fernandez\unskip,\iUCSC}
\mbox{M.J. Fero\unskip,\iMIT}
\mbox{K. Flood\unskip,\iMASS}
\mbox{R. Frey\unskip,\iOREG}
\mbox{J. Gifford\unskip,\iUVIC}
\mbox{T. Gillman\unskip,\iRAL}
\mbox{G. Gladding\unskip,\iILLI}
\mbox{S. Gonzalez\unskip,\iMIT}
\mbox{E.R. Goodman\unskip,\iCOLO}
\mbox{E.L. Hart\unskip,\iTENN}
\mbox{J.L. Harton\unskip,\iCSU}
\mbox{K. Hasuko\unskip,\iTOHO}
\mbox{S.J. Hedges\unskip,\iBU}
\mbox{S.S. Hertzbach\unskip,\iMASS}
\mbox{M.D. Hildreth\unskip,\iSLAC}
\mbox{J. Huber\unskip,\iOREG}
\mbox{M.E. Huffer\unskip,\iSLAC}
\mbox{E.W. Hughes\unskip,\iSLAC}
\mbox{X. Huynh\unskip,\iSLAC}
\mbox{H. Hwang\unskip,\iOREG}
\mbox{M. Iwasaki\unskip,\iOREG}
\mbox{D.J. Jackson\unskip,\iRAL}
\mbox{P. Jacques\unskip,\iRUTG}
\mbox{J.A. Jaros\unskip,\iSLAC}
\mbox{Z.Y. Jiang\unskip,\iSLAC}
\mbox{A.S. Johnson\unskip,\iSLAC}
\mbox{J.R. Johnson\unskip,\iWISC}
\mbox{R.A. Johnson\unskip,\iCINC}
\mbox{T. Junk\unskip,\iSLAC}
\mbox{R. Kajikawa\unskip,\iNAGO}
\mbox{M. Kalelkar\unskip,\iRUTG}
\mbox{Y. Kamyshkov\unskip,\iTENN}
\mbox{H.J. Kang\unskip,\iRUTG}
\mbox{I. Karliner\unskip,\iILLI}
\mbox{H. Kawahara\unskip,\iSLAC}
\mbox{Y.D. Kim\unskip,\iSOGA}
\mbox{M.E. King\unskip,\iSLAC}
\mbox{R. King\unskip,\iSLAC}
\mbox{R.R. Kofler\unskip,\iMASS}
\mbox{N.M. Krishna\unskip,\iCOLO}
\mbox{R.S. Kroeger\unskip,\iMISSI}
\mbox{M. Langston\unskip,\iOREG}
\mbox{A. Lath\unskip,\iMIT}
\mbox{D.W.G. Leith\unskip,\iSLAC}
\mbox{V. Lia\unskip,\iMIT}
\mbox{C.Lin\unskip,\iMASS}
\mbox{M.X. Liu\unskip,\iYALE}
\mbox{X. Liu\unskip,\iUCSC}
\mbox{M. Loreti\unskip,\iPADO}
\mbox{A. Lu\unskip,\iUCSB}
\mbox{H.L. Lynch\unskip,\iSLAC}
\mbox{J. Ma\unskip,\iWASH}
\mbox{M. Mahjouri\unskip,\iMIT}
\mbox{G. Mancinelli\unskip,\iRUTG}
\mbox{S. Manly\unskip,\iYALE}
\mbox{G. Mantovani\unskip,\iPERU}
\mbox{T.W. Markiewicz\unskip,\iSLAC}
\mbox{T. Maruyama\unskip,\iSLAC}
\mbox{H. Masuda\unskip,\iSLAC}
\mbox{E. Mazzucato\unskip,\iFERR}
\mbox{A.K. McKemey\unskip,\iBRUN}
\mbox{B.T. Meadows\unskip,\iCINC}
\mbox{G. Menegatti\unskip,\iFERR}
\mbox{R. Messner\unskip,\iSLAC}
\mbox{P.M. Mockett\unskip,\iWASH}
\mbox{K.C. Moffeit\unskip,\iSLAC}
\mbox{T.B. Moore\unskip,\iYALE}
\mbox{M.Morii\unskip,\iSLAC}
\mbox{D. Muller\unskip,\iSLAC}
\mbox{V. Murzin\unskip,\iMOSCOW}
\mbox{T. Nagamine\unskip,\iTOHO}
\mbox{S. Narita\unskip,\iTOHO}
\mbox{U. Nauenberg\unskip,\iCOLO}
\mbox{H. Neal\unskip,\iSLAC}
\mbox{M. Nussbaum\unskip,\iCINC}
\mbox{N. Oishi\unskip,\iNAGO}
\mbox{D. Onoprienko\unskip,\iTENN}
\mbox{L.S. Osborne\unskip,\iMIT}
\mbox{R.S. Panvini\unskip,\iVAND}
\mbox{C.H. Park\unskip,\iSOONG}
\mbox{T.J. Pavel\unskip,\iSLAC}
\mbox{I. Peruzzi\unskip,\iFRAS}
\mbox{M. Piccolo\unskip,\iFRAS}
\mbox{L. Piemontese\unskip,\iFERR}
\mbox{K.T. Pitts\unskip,\iOREG}
\mbox{R.J. Plano\unskip,\iRUTG}
\mbox{R. Prepost\unskip,\iWISC}
\mbox{C.Y. Prescott\unskip,\iSLAC}
\mbox{G.D. Punkar\unskip,\iSLAC}
\mbox{J. Quigley\unskip,\iMIT}
\mbox{B.N. Ratcliff\unskip,\iSLAC}
\mbox{T.W. Reeves\unskip,\iVAND}
\mbox{J. Reidy\unskip,\iMISSI}
\mbox{P.L. Reinertsen\unskip,\iUCSC}
\mbox{P.E. Rensing\unskip,\iSLAC}
\mbox{L.S. Rochester\unskip,\iSLAC}
\mbox{P.C. Rowson\unskip,\iCOLU}
\mbox{J.J. Russell\unskip,\iSLAC}
\mbox{O.H. Saxton\unskip,\iSLAC}
\mbox{T. Schalk\unskip,\iUCSC}
\mbox{R.H. Schindler\unskip,\iSLAC}
\mbox{B.A. Schumm\unskip,\iUCSC}
\mbox{J. Schwiening\unskip,\iSLAC}
\mbox{S. Sen\unskip,\iYALE}
\mbox{V.V. Serbo\unskip,\iSLAC}
\mbox{M.H. Shaevitz\unskip,\iCOLU}
\mbox{J.T. Shank\unskip,\iBU}
\mbox{G. Shapiro\unskip,\iLBL}
\mbox{D.J. Sherden\unskip,\iSLAC}
\mbox{K.D. Shmakov\unskip,\iTENN}
\mbox{C. Simopoulos\unskip,\iSLAC}
\mbox{N.B. Sinev\unskip,\iOREG}
\mbox{S.R. Smith\unskip,\iSLAC}
\mbox{M.B. Smy\unskip,\iCSU}
\mbox{J.A. Snyder\unskip,\iYALE}
\mbox{H. Staengle\unskip,\iCSU}
\mbox{A. Stahl\unskip,\iSLAC}
\mbox{P. Stamer\unskip,\iRUTG}
\mbox{H. Steiner\unskip,\iLBL}
\mbox{R. Steiner\unskip,\iADEL}
\mbox{M.G. Strauss\unskip,\iMASS}
\mbox{D. Su\unskip,\iSLAC}
\mbox{F. Suekane\unskip,\iTOHO}
\mbox{A. Sugiyama\unskip,\iNAGO}
\mbox{S. Suzuki\unskip,\iNAGO}
\mbox{M. Swartz\unskip,\iJHU}
\mbox{A. Szumilo\unskip,\iWASH}
\mbox{T. Takahashi\unskip,\iSLAC}
\mbox{F.E. Taylor\unskip,\iMIT}
\mbox{J. Thom\unskip,\iSLAC}
\mbox{E. Torrence\unskip,\iMIT}
\mbox{N.K. Toumbas\unskip,\iSLAC}
\mbox{T. Usher\unskip,\iSLAC}
\mbox{C. Vannini\unskip,\iPISA}
\mbox{J. Va'vra\unskip,\iSLAC}
\mbox{E. Vella\unskip,\iSLAC}
\mbox{J.P. Venuti\unskip,\iVAND}
\mbox{R. Verdier\unskip,\iMIT}
\mbox{P.G. Verdini\unskip,\iPISA}
\mbox{D.L. Wagner\unskip,\iCOLO}
\mbox{S.R. Wagner\unskip,\iSLAC}
\mbox{A.P. Waite\unskip,\iSLAC}
\mbox{S. Walston\unskip,\iOREG}
\mbox{S.J. Watts\unskip,\iBRUN}
\mbox{A.W. Weidemann\unskip,\iTENN}
\mbox{E. R. Weiss\unskip,\iWASH}
\mbox{J.S. Whitaker\unskip,\iBU}
\mbox{S.L. White\unskip,\iTENN}
\mbox{F.J. Wickens\unskip,\iRAL}
\mbox{B. Williams\unskip,\iCOLO}
\mbox{D.C. Williams\unskip,\iMIT}
\mbox{S.H. Williams\unskip,\iSLAC}
\mbox{S. Willocq\unskip,\iMASS}
\mbox{R.J. Wilson\unskip,\iCSU}
\mbox{W.J. Wisniewski\unskip,\iSLAC}
\mbox{J. L. Wittlin\unskip,\iMASS}
\mbox{M. Woods\unskip,\iSLAC}
\mbox{G.B. Word\unskip,\iVAND}
\mbox{T.R. Wright\unskip,\iWISC}
\mbox{J. Wyss\unskip,\iPADO}
\mbox{R.K. Yamamoto\unskip,\iMIT}
\mbox{J.M. Yamartino\unskip,\iMIT}
\mbox{X. Yang\unskip,\iOREG}
\mbox{J. Yashima\unskip,\iTOHO}
\mbox{S.J. Yellin\unskip,\iUCSB}
\mbox{C.C. Young\unskip,\iSLAC}
\mbox{H. Yuta\unskip,\iAOMORI}
\mbox{G. Zapalac\unskip,\iWISC}
\mbox{R.W. Zdarko\unskip,\iSLAC}
\mbox{J. Zhou\unskip.\iOREG}

\it
  \vskip \baselineskip                   
  \centerline{(The SLD Collaboration)}   
  \vskip \baselineskip        
  \baselineskip=.75\baselineskip   
\iADEL
  Adelphi University, Garden City, New York 11530, \break
\iAOMORI
  Aomori University, Aomori , 030 Japan, \break
\iBOLO
  INFN Sezione di Bologna, I-40126, Bologna, Italy, \break
\iBRI
  University of Bristol, Bristol, U.K., \break
\iBRUN
  Brunel University, Uxbridge, Middlesex, UB8 3PH United Kingdom, \break
\iBU
  Boston University, Boston, Massachusetts 02215, \break
\iCINC
  University of Cincinnati, Cincinnati, Ohio 45221, \break
\iCOLO
  University of Colorado, Boulder, Colorado 80309, \break
\iCOLU
  Columbia University, New York, New York 10533, \break
\iCSU
  Colorado State University, Ft. Collins, Colorado 80523, \break
\iFERR
  INFN Sezione di Ferrara and Universita di Ferrara, I-44100 Ferrara, Italy, \break
\iFRAS
  INFN Lab. Nazionali di Frascati, I-00044 Frascati, Italy, \break
\iILLI
  University of Illinois, Urbana, Illinois 61801, \break
\iJHU
  Johns Hopkins University,  Baltimore, Maryland 21218-2686, \break
\iLBL
  Lawrence Berkeley Laboratory, University of California, Berkeley, California 94720, \break
\iLTU
  Louisiana Technical University, Ruston,Louisiana 71272, \break
\iMASS
  University of Massachusetts, Amherst, Massachusetts 01003, \break
\iMISSI
  University of Mississippi, University, Mississippi 38677, \break
\iMIT
  Massachusetts Institute of Technology, Cambridge, Massachusetts 02139, \break
\iMOSCOW
  Institute of Nuclear Physics, Moscow State University, 119899, Moscow Russia, \break
\iNAGO
  Nagoya University, Chikusa-ku, Nagoya, 464 Japan, \break
\iOREG
  University of Oregon, Eugene, Oregon 97403, \break
\iOXF
  Oxford University, Oxford, OX1 3RH, United Kingdom, \break
\iPADO
  INFN Sezione di Padova and Universita di Padova I-35100, Padova, Italy, \break
\iPERU
  INFN Sezione di Perugia and Universita di Perugia, I-06100 Perugia, Italy, \break
\iPISA
  INFN Sezione di Pisa and Universita di Pisa, I-56010 Pisa, Italy, \break
\iRAL
  Rutherford Appleton Laboratory, Chilton, Didcot, Oxon OX11 0QX United Kingdom, \break
\iRUTG
  Rutgers University, Piscataway, New Jersey 08855, \break
\iSLAC
  Stanford Linear Accelerator Center, Stanford University, Stanford, California 94309, \break
\iSOGA
  Sogang University, Seoul, Korea, \break
\iSOONG
  Soongsil University, Seoul, Korea 156-743, \break
\iTENN
  University of Tennessee, Knoxville, Tennessee 37996, \break
\iTOHO
  Tohoku University, Sendai 980, Japan, \break
\iUCSB
  University of California at Santa Barbara, Santa Barbara, California 93106, \break
\iUCSC
  University of California at Santa Cruz, Santa Cruz, California 95064, \break
\iUVIC
  University of Victoria, Victoria, British Columbia, Canada V8W 3P6, \break
\iVAND
  Vanderbilt University, Nashville,Tennessee 37235, \break
\iWASH
  University of Washington, Seattle, Washington 98105, \break
\iWISC
  University of Wisconsin, Madison,Wisconsin 53706, \break
\iYALE
  Yale University, New Haven, Connecticut 06511. \break

\rm
%

\end{center}

\begin{table}
\caption{Relative systematic errors on the 1996-98 vertex charge 
measurement of $A_b$.}
\begin{center}
\begin{tabular}{lcc}
{\bf Error Source} & {\bf Variation} & $\delta A_b/A_b$ \\ \hline
 & & \\
\multicolumn{3}{l}{\it \underline{Self-Calibration}} \\ 
$p_b$ statistics & $\pm$1$\sigma$ & +3.4/-3.2\% \\
Self-cal bias & MC Statistics & 1.0\% \\
$\cos\theta$ shape of $p_b$ & MC Shape $vs$ Flat & 0.2\% \\
Light Flavor  & $\pm 50\%$ of correction & 0.1\% \\
 & & \\
\multicolumn{3}{l}{\it \underline{Analysis}} \\
Tag Composition & $\Pi_b \pm \delta\Pi_b$  & 0.9\% \\
Detector Modeling & Tracking eff.  & 1.0\% \\
 &and resolution& \\
 &corrections on/off& \\
Beam Polarization & $\pm$0.8\% & 0.8\% \\
QCD & $x_{QCD}$, $\alpha_s\pm 0.007$,& 0.8\% \\
    &$ 2^{nd}$ order terms &\\
Gluon Splitting & $\pm$100\% of JETSET& 0.2\% \\
$A_c$ & $0.67\pm 0.05$ & $<$0.1\% \\
$A_{bckg}$ & $0\pm 0.50$ & $<$0.1\% \\
 \hline
{\bf Total} & & {\bf +4.0/-3.8\%}
\end{tabular}
\end{center}
\label{SYSERR}
\end{table}

\end{document}